\documentclass[12pt]{article}
\usepackage[english]{babel}
\usepackage{hyperref,hypcap,ae,tikz}
\usepackage{graphicx}
\usepackage{bm}
\usepackage{amsmath}
\usepackage{amssymb}
\usepackage{mathrsfs}
\usepackage{mathtools}
\usepackage{amsfonts}
\usepackage[T1]{fontenc}
\usepackage{textcomp}
\usepackage{xcolor}
\usepackage{lmodern}
\usepackage[caption=false]{subfig}
\usepackage{booktabs}
\usepackage{cite}
\usepackage[section]{placeins}
\usepackage{authblk}
\hypersetup{
	colorlinks=true,
	citecolor=blue,
    linkcolor=blue,
    urlcolor=blue,
	}
	
\date{}	
\begin{document}

\title{{\bf Complexity and structure scalars of Type II matter fields}}

\author[1]{Samarjit Chakraborty \thanks{\href{mailto:samarjitxxx@gmail.com}{samarjitxxx@gmail.com}}}
\author[1]{Rituparno Goswami \thanks{\href{mailto:Goswami@ukzn.ac.za}{Goswami@ukzn.ac.za}}}
\author[1]{Sunil D. Maharaj \thanks{\href{mailto:MAHARAJ@ukzn.ac.za}{MAHARAJ@ukzn.ac.za}}}
\affil[1]{Astrophysics Research Centre, School of Mathematics, Statistics and Computer Science, University of KwaZulu-Natal, Private Bag X54001, Durban 4000, South Africa.}

\maketitle
\section*{Abstract}
A general semi-tetrad covariant approach is adopted to analyse the structure scalars of a Type II fluid in generalized Vaidya spacetime. The relationship between the $1+1+2$ covariant quantities and the structure scalars are obtained. We calculate the complexity factor in terms of the Misner-Sharp mass and the matter variables to obtain a non-trivial class of spacetimes with vanishing complexity. Also the Vaidya spacetime with pure Type II matter field has negative complexity. The differences between the complexity of Type I and Type II matter fields are highlighted. We compute the propagation and evolution equations of the structure scalars, showcasing their interdependency through the kinematical variables. The causal wave equation of the Gaussian curvature of the 2-shell and its dependence on the structure scalars are also studied.

\section{Introduction}
The notion of complexity is an intuitive concept which simply indicates how complicated a given system is. It is obvious to think that the underlying mechanisms are responsible for this phenomenon. Therefore, a simple system has a very low complexity, with a minimal underlying dynamics. As the system becomes more complicated the complexity rises. Conceptually, the idea is simple. However, when we try to quantify complexity, an unique definition is difficult to obtain. A recently proposed measure of complexity by Herrera et al \cite{H1}
defines complexity (classically) via one of the structure scalars, obtained through the decomposition of the Riemann tensor using a preferred timelike observer \cite{H0}. The complexity factor $ Y_{TF} $ (trace-free component of the projected tensor) is related to the electric part of the Weyl tensor and the pressure anisotropy. This is very useful because the electric part of Weyl is related to the density inhomogeneity and other dissipative variables. As we only consider the spherically symmetric case, the magnetic part of the Weyl tensor is zero. Therefore, the complexity factor encompasses the complete tidal contribution. 

However, what happens to the complexity in a null radiation filled spacetime (a matter field very common in astrophysics)? The nature of the complexity factor of such a matter field is of great importance. It helps us to understand the null dust field and the transition of complexity at the stellar boundary. Ideally we can think of this field as a mix of null radiation and matter. To be specific, we are concerned about the complexity of Type II matter fields (null dust radiation plus matter). They appear in non-static spacetimes with radiation, like a radiating black hole (BH) or the exterior of a radiating star. Therefore, to understand the complexity of the entire stellar structure also (e.g., Type I matter in the interior and Type II radiation field in the exterior.), the study of Type II fields become necessary. Hence, we consider the generalized Vaidya class of spacetimes as the appropriate background. These spacetimes are important tools in many areas of investigation, namely, the study of cosmic censorship conjecture (CCC), formation of naked singularities, BH evaporation, Hawking radiation etc. The recent work by Dadhich and Goswami \cite{DaGo} on the relationship between classical Vaidya radiation and quantum Hawking radiation in an accreting BH again highlights the importance of Type II matter fields. The authors show that preserving the degeneracy between the global event horizon and the locally defined null marginally outer trapped surface (MOTS) requires an infalling Type I matter field to transition to Type II as it approaches the horizon, resulting in an outward directed heat flux. Additionally, during extreme collapse situation, Type II matter becomes more significant than Type I. Therefore, the study of Type II matter field complexity has deep significance. 

We now briefly mention some of the important previous works in this field.

\subsection{Previous works on complexity}
We adopt the definition of complexity proposed by Herrera in \cite{H1}, where the complexity of the self gravitating spherically symmetric static systems are introduced. This work was further extended to dynamical systems by Herrera et al. \cite{H2}. Subsequently, Herrera and collaborators studied axially symmetric static cases in \cite{H3}. Further studies in cylindrical symmetry were done by Sharif et al. in \cite{S1,S2}. A significant amount of work on this formulation of complexity can be found in \cite{A1,A2,A3,A4,A5}. Recently,  complexity of static fluid systems in hyperbolic symmetry were examined in \cite{Hy1,Hy2,Hy3}. Moreover, many other works consider this definition of complexity in modified theories of gravity \cite{M1,M2,M3,M4,M5,M6,M7}. For a detailed review of the works (on this complexity prescription) can be found in \cite{B} and the references therein. In recent years, Naseer, Sharif and others have extensively worked on complexity (of Type I fluids) in the context of modified theories of gravity which includes $f(R)$, $f(R,T)$, Rastall gravity etc. in \cite{TN1,TN2,TN3,TN4,TN5} and found different solutions for vanishing complexity. These works and many others in modified gravity indicate the wider applicability of Herrera's complexity factor.

\subsection{This paper}
In view of the previously mentioned works on Type I fluids, the study of Type II matter field complexity in general relativity (GR) becomes important (given the wide application in astrophysics). To the best of our knowledge, no detailed study on the complexity of null fluids are available in the literature. Therefore, this paper primarily deals with the following problem.\\
\textbf{Key question:} Is the complexity factor of Type II matter field different from the Type I complexity?

To answer this question we first adapt the formulation of complexity proposed in \cite{H1}, into 1+1+2 covariant description. This is necessary, as the structure scalars are defined using timelike observers, whereas we will be dealing with null fields (Type II). We have refrained ourselves from describing the formalism in detail as it is widely studied with many literatures available. The formalism gives us a set of covariant scalars which helps us to understand the system (and track the physical processes) in much more clarity. The relevant governing equations also simplify to a great degree in LRS II spacetimes, as it helps us to bypass the spherically symmetric 2-shell (trivial) and minimizes the complexity of calculations in comparison to the standard GR or full tetrad Newman-Penrose formalism. Hence a canonical 1+1+2 covariant description becomes helpful.

As mentioned before, we are interested in physically meaningful systems. In astrophysical context, Type II matter fields usually occur in radiating star systems (exterior spacetime) and dynamic compact objects. This is described by the Vaidya class of spacetimes with null dust. Therefore, the generalized Vaidya spacetime (pure null dust mixed with matter) is considered \cite{WW,DG} as our relevant Type II matter field. We obtain the covariant expression of the structure scalars and show how the complexity factor behaves differently for a Type II matter in pure Vaidya spacetime.

Apart from the above problem we also investigated the following aspects:

\begin{itemize}
\item How to express the structure scalars in 1+1+2 covariant description?
\item What is the relationship among the structure scalars for a Type II matter field?
\item What are the propagation and evolution equations of these scalars in the 1+1+2 covariant description? 
\item What is the influence of these scalars on the Gaussian curvature and Misner-Sharp mass on the spherical 2-shell?
\end{itemize}

We first introduce the general semi-tetrad formulation for LRS II spacetimes, which contains the subclass of spherical symmetry. Then we briefly discuss the covariant decomposition of the Riemann tensor and how the structure scalars are defined. We express the structure scalars and curvature tensors in terms of 1+1+2 covariant quantities. Finally, we consider the generalized Vaidya spacetime and obtain the complexity factor. Different subclass solutions are discussed belonging to varying complexities. The evolution and propagation equations of these scalars are also obtained. In the next portion we discuss how these scalars influence the wave equation of the Gaussian curvature of the 2-shell.  Subsequently, we summarize and conclude our study in the last section. 
\section{Covariant description of LRS II spacetimes}
In this section we briefly summarize the semi-tetrad 1+1+2 covariant description of LRS II spacetimes, which contains the spherically symmetric spacetimes as a subclass. These spacetimes are non-rotating and contain a continuous non-trivial isotropy group of spatial rotations at every point. For more details please refer to \cite{E1,E2,E3,E4,E5}. 
Hence, we have a covariantly defined preferred spatial direction at each point. Now, by choosing the timelike unit vector $u^a$ (usually defined along the fluid flow lines) and the orthogonal unit vector along the preferred spatial direction $e^a$, we can decompose the spacetime as \cite{E6}
\begin{equation}\label{decomp}
g_{ab}=-u_au_b+e_ae_b+N_{ab},
\end{equation}
where $N_{ab}$ is the projection tensor on the spherical 2-shells and is defined as $ N_{ab}=h_{ab}-e_{a}e_{b} $. Here $ h_{ab} $ is the projection tensor orthogonal to $ u^{a} $ and spans the $3$-space.

Therefore, in the $1+3$ description this gives rise to two kinds of derivatives: 
\begin{itemize}
\item \textbf{The dot derivative}: This is the covariant time derivative along the observers' worldlines . Therefore, for any tensor  $ S^{a...b}{}_{c...d}$ it is defined as 
$\dot{S}^{a...b}{}_{c...d}\, \equiv u^{e} \nabla_{e} {S}^{a...b}{}_{c...d} $.
\item \textbf{The \textit{D} derivative}: This spatial derivative is defined by orthogonally projecting the covariant derivative onto the 3-space using the projection tensor $ h_{ab} $. Hence we have 
$ D_{e}S^{a...b}{}_{c...d}{} \equiv h^a{}_f
h^p{}_c...h^b{}_g h^q{}_d h^r{}_e \nabla_{r} {S}^{f...g}{}_{p...q}$.
\end{itemize} 
Further, in the $1+1+2$ formulation the preferred spatial vector $e^a$ introduces two new derivatives by splitting the 3-space, for any 3-tensor $ \psi_{a...b}{}^{c...d} $ :
\begin{itemize}
\item \textbf{The hat derivative}: This is the spatial derivative along the vector $e^a$ defined as $\hat{\psi}_{a...b}{}^{c...d} \equiv e^{f}D_{f}\psi_{a...b}{}^{c...d}$.
\item \textbf{The delta derivative}: It is the projected spatial derivative on the 2-sheet by the projection tensor $N_a^{~b}$ on all the free indices. Hence 
$\delta_f\psi_{a...b}{}^{c...d} \equiv N_{a}{}^{f}...N_{b}{}^gN_{h}{}^{c}...
N_{i}{}^{d}N_f{}^jD_j\psi_{f...g}{}^{i...j}$.
\end{itemize}

As a result, we get the geometrical quantities for the chosen timelike congruence $ u^{a} $. These are the expansion scalar $\Theta$, acceleration 3-vector $\dot{u}^a$ and the shear 3-tensor $\sigma_{ab}$.
It also uniquely defines the electric part of the Weyl tensor (which is responsible for tidal forces and inhomogeneity), as $E_{ab}= C_{acbd}u^cu^d$. Moreover, the magnetic part of the Weyl tensor (due to  rotation or time varying  spacetime) $H_{ab}=C^*_{acbd}u^cu^d$, vanishes identically \cite{E7}. This timelike congruence $u^a$ also splits the energy momentum tensor of the matter field (EMT) into the energy density $\rho$, isotropic pressure $p$, heat flux 3-vector $q^a$ and the anisotropic stress 3-tensor $\pi_{ab}$.

In this class of spacetimes the only non-vanishing geometrical quantity related to the preferred spacelike congruence $ e^{a} $ is the volume expansion $\phi=\delta_ae^a$, as other geometrical quantities due to this congruence vanish identically in spherical symmetry.

Further, we can extract a set of covariant scalars from the above mentioned 3-vectors and 3-tensors as $\mathcal{A}=\dot{u}^ae_a$, $\Sigma=\sigma_{ab}e^ae^b$, $\mathcal{E}=E_{ab}e^ae^b$, $Q=q^ae_a$ and $\Pi=\pi_{ab}e^ae^b$. Finally, we get the set of scalars that fully describes the spherically symmetric class of spacetimes
\begin{equation}\label{set1}
\mathcal{D}\equiv\left\{\Theta, \mathcal{A}, \Sigma, \mathcal{E}, \phi, \rho, p, \Pi, Q\right\}.
\end{equation}
The obtained geometrical and thermodynamical scalars (from EMT), with their directional derivatives along $u^a$ (denoted by a dot) and $e^a$ (denoted by a hat) completely specify the Ricci identities and the Bianchi identities. Therefore, they can completely specify the dynamics of the spacetime. Another useful geometrically defined quantity is the Misner-Sharp mass $ \mathcal{M} $ (the amount of mass enclosed within the spherical $2$-shell at a given instant of time). In terms of the covariant scalars it is given as \cite{E8,E9}
\begin{equation}\label{Mass}
\mathcal{M} = \frac{1}{2K^{\frac{3}{2}}}\left(\frac{1}{3}\rho - \mathcal{E} - \frac{1}{2}\Pi\right).
\end{equation}
Hence, in vacuum spacetimes the mass is entirely due to the electric part of Weyl tensor $\cal E$. Here $K$ is the Gaussian curvature of the spherical shells, which provides a geometric notion of the area radius of the spherical shells $\mathcal{R}$, and is defined as the inverse square of $\mathcal{R}$. The Gaussian curvature is also related to the obtained covariant scalars in Eq.(\ref{set1}) as  
\begin{equation}\label{gauss}
K \equiv\frac{1}{\mathcal{R}^2}=\frac{1}{3}\rho-\mathcal{E}-\frac{1}{2}\Pi +\frac{1}{4}\phi^2
-\left(\frac{1}{3}\Theta-\frac{1}{2}\Sigma\right)^2\, .
\end{equation}
Other important relations are the directional derivatives of $\mathcal{M}$ along $u^a$ and $e^a$ in terms of the covariant scalars  
\begin{eqnarray}
\label{MHat}
\hat{\mathcal{M}} &=& \frac{1}{4K^{\frac{3}{2}}}\left[\phi\rho - \left(\Sigma-\frac{2}{3}\Theta\right)Q\right],\\
\label{MDot}
\dot{\mathcal{M}} &=&  \frac{1}{4K^{\frac{3}{2}}}\left[\left(\Sigma - \frac{2}{3}\Theta\right)\left(p + \Pi\right) -\phi Q\right].
\end{eqnarray}
Moreover, the directional derivatives of any scalar $ \mathcal{O} $ in LRS II spacetime have to satisfy the following commutation relation
\begin{equation}\label{Comm}
\hat{\dot{\mathcal{O}}}-\dot{\hat{\mathcal{O}}}=-\mathcal{A}\dot{\mathcal{O}}+\left(\dfrac{1}{3}\Theta+\Sigma\right)\hat{\mathcal{O}}.
\end{equation}
Due to this $1+1+2$ decomposition we can write the EMT of a general matter field in the following manner
\begin{equation}\label{EM1}
T_{ab}=\rho u_au_b + (p+\Pi)e_ae_b +2Qu_{(a}e_{b)}+\left(p-\frac12\Pi\right)N_{ab}\;.
\end{equation}
Now in the $[u,e]$ plane (the spherical 2-shells are trivial), the discriminant ($\Delta$) of the quadratic equations for the eigenvalues of the above EMT tensor (\ref{EM1}) takes the form
\begin{equation}\label{eigen1}
\Delta= (\rho+p+\Pi)^2-4Q^2\;.
\end{equation}
Therefore, if $\vert Q\vert < \frac12(\rho+p+\Pi)$, there exists two distinct eigenvalues corresponding to one timelike and one spacelike eigenvectors on this plane, making the matter field necessarily of Type I (timelike matter field). 

However, when we have the following relation
\begin{equation}\label{eigen2}
\vert Q\vert =\frac12(\rho+p+\Pi),
\end{equation}
it makes both the eigenvalues equal (with double degenerate null eigenvectors) and fixes the value of the heat flux $Q$. Therefore, the resulting matter field is always of Type II (null). Please note that in our subsequent calculations we take the Einstein's gravitational constant $ \kappa=1 $.

\section{Orthogonal decomposition of Riemann tensor: structure scalars and complexity}

The orthogonal splitting of the Riemann tensor was first studied in \cite{SP1} and since then it has been used in many places.
Define the left, right and double dual of Riemann tensor in the standard 
fashion
$$
\ ^{*}R_{abcd}\equiv\frac{1}{2}\eta_{abpq}R^{pq}_{\ \ cd},\ 
R^{*}_{abcd}\equiv\frac{1}{2}\eta_{pqcd}R_{ab}^{\ \ pq},\ 
\ ^{*}R^*_{abcd}=\frac{1}{2}\eta_{ab}^{\ \ pq}R^*_{pqcd}.
$$
Next we introduce the following spatial tensors \cite{SP2}
\begin{equation}
Y_{ac}\equiv R_{abcd}u^bu^d,\ Z_{ac}\equiv \ ^{*}R_{abcd}u^bu^d,\ 
X_{ac}\equiv\ ^{*}R^*_{abcd}u^bu^d,
\label{riemann-parts}
\end{equation}
where the symmetries of Riemann tensor entail the following properties
\begin{equation}
X_{(ab)}=X_{ab},\ Y_{(ab)}=Y_{ab},\ Z^a_{\ a}=0.
\label{riemann-parts-prop}
\end{equation}
These tensors contain all the information in the Riemann tensor as is easily checked by a simple count of their total number of independent components. They also enable
us to find the orthogonal splitting (1+3) of the Riemann tensor which reads
\begin{eqnarray}
 R_{{abcd}}^{{    }}=2u_cu_{[a}Y_{{b]d}}^{{  }}+2h_{{a[d}}^{{  }} X_{{c]b}}^{{ }}
+2u_du_{[b} Y_{{a]c}}^{{  }}+2u_{[d} Z_{{\ c]}}^{{e }}\epsilon_{{abe}}^{{   }}
+2u_{[b} Z_{{\ a]}}^{{e }}\epsilon_{{cde}}+\nonumber\\
+h_{{bd}}^{{  }}\left(h_{{ac}}^{{  }} 
X_{{\  e}}^{{e }}-X_{{ac}}^{{ }}\right)+h_{{bc}}^{{  }}
\left(X_{{ad}}^{{ }}-h_{{ad}}^{{  }} X_{{\   e}}^{{e }}\right),
\label{riemann-split}
\end{eqnarray}
where 
\begin{equation}
\eta_{abcd}=-u_a\epsilon_{bcd}+u_b\epsilon_{acd}
-u_c\epsilon_{abd}+u_d\epsilon_{abc}.
\end{equation}
Here $\epsilon_{abc}$ is the {\em spatial volume element} 
and is defined by $\epsilon_{abc}\equiv u^d\eta_{dabc}, \epsilon_{abc}u^{c}=0.$ From the above expression we get the orthogonal splitting of 
the Ricci tensor as
\begin{equation}
R_{ac}=Z^{db} \epsilon_{cdb}u_a+u_cY_{\ d}^{d } u_a-X_{ac}-Y_{ac}
+u_c Z^{db} \epsilon_{adb}+h_{ac} X_{\ d}^{d}.
\label{ricci-split}
\end{equation}
These second rank tensors (\ref{riemann-parts}) are decomposed into their trace and trace-free parts to obtain the `structure scalars'.

In order to get a canonical covariant description in semi-tetrad formalism, we first adapt the Herrera's complexity prescription in 1+1+2 formulation. We will use the general form of EMT and decompose it accordingly. We begin our analysis with the matter energy momentum tensor $T_{ab}$ of the fluid distribution involving the energy density $\mu$, radial pressure $P_r$ , tangential pressure $P_{\perp}$, heat flux  $q^{a}$,  four velocity of the fluid $V^{a}$, and the unit four vector along the radial direction $\chi^{a}$. These quantities also satisfy
\begin{eqnarray}
V^{a}V_{a}=-1, \;\; V^{a}q_{a}=0, \;\; \chi^{a}\chi_{a}=1,\;\;
\chi^{a}V_{a}=0.
\end{eqnarray}
Therefore, the energy momentum tensor in the canonical form is the following
\begin{equation}
T_{ab} = {\mu} V_a V_b + P h_{ab} + \Pi_{ab} +
q \left(V_a \chi_b + \chi_a V_b\right), \label{Tab}
\end{equation}
with
$$ P=\frac{P_{r}+2P_{\bot}}{3}, \qquad h_{ab}=g_{ab}+V_a V_b,$$

$$\Pi_{ab}=\tilde{\Pi}\left(\chi_a \chi_b - \frac{1}{3} h_{ab}\right), \qquad \tilde{\Pi}=P_{r}-P_{\bot}.$$

From the previously discussed orthogonal decomposition of Riemann tensor and the matter stress-energy tensor we can split the curvature tensors in terms of matter variables. Subsequently,  the tensors $Y_{ab}$ and $X_{ab}$ may be expressed (trace and trace-free parts) in terms of four scalar functions $Y_T, Y_{TF}, X_T, X_{TF}$ (structure scalars) and the scalar $ Z $ is constructed from the self contraction of the tensor $ Z_{ab} $ by 
\begin{eqnarray}
Y_{ab}=\frac{1}{3}Y_T h_{ab}+Y_{TF}\left(\chi_{a} \chi_{b}-\frac{1}{3}h_{ab}\right),\label{electric'}
\\
X_{ab}=\frac{1}{3}X_T h_{ab}+X_{TF}\left(\chi_{a} \chi_{b}-\frac{1}{3}h_{ab}\right),\label{magnetic'}
\\
Z=\sqrt{Z_{ab}Z^{ab}}.\label{fluxterm}
\end{eqnarray}
In spherical symmetry the magnetic part of the Weyl tensor vanishes and therefore the electric part of the Weyl tensor, may be written as
\begin{equation}
E_{ab}=\mathbb{E} \left(\chi_a \chi_b-\frac{1}{3}h_{ab}\right).
\label{52}
\end{equation}
After using the field equations we can obtain the structure scalars.
For a detailed derivation please see \cite{H1,H0,H2,B,SP2}. 

Now, we can always take the corresponding timelike vector $ V^{a} $ same as $ u^{a} $ and the orthogonal spacelike vector $ \chi^{a} $ along $ e^{a} $. Consequently, we can split the projection tensor $ h_{ab} $ along $ e^{a} $ and the sheet projection tensor $ N_{ab} $ to get the following relations
\begin{equation}
\rho=\mu,\,Q=q,\,p=P,\, \Pi=\dfrac{2\tilde{\Pi}}{3},\,\, \mathcal{E}=\dfrac{2\mathbb{E}}{3},
\end{equation}
where $\{\rho, Q, p, \Pi, \mathcal{E}\}$ are the covariant scalars in 1+1+2 decomposition. Further, in LRS II spacetimes the matter stress-energy tensor can be rewritten as Eq.(\ref{EM1}). Finally, in terms of 1+1+2 variables the structure scalars for LRS II spacetime can be expressed as
\begin{eqnarray}\label{structure}
Y_T=(\rho+3p)/2, \qquad
Y_{TF}=\dfrac{3}{2}\left(\mathcal{E}-\Pi/2\right) ,\nonumber
\\
X_T= \rho , \qquad
X_{TF}=-\dfrac{3}{2}\left(\mathcal{E}+ \Pi/2\right),\nonumber
\\
\centering Z= Q/\sqrt{2}.
\end{eqnarray}
Using Eq.(\ref{Mass}) we can also express the Misner-sharp mass (per volume) in terms of the structure scalars (magnetic part of the Riemann tensor) in the following manner
\begin{equation}
\dfrac{1}{2}X_{T}+X_{TF}=\dfrac{3\mathcal{M}}{\mathcal{R}^3}.
\end{equation}
The magnetic and electric parts of the Riemann tensor in 1+1+2 decomposition can now be written as
\begin{align}
X_{ab}=\dfrac{2\mathcal{M}}{{\mathcal{R}}^3}e_ae_b+\dfrac{1}{2}\left(\rho-\dfrac{2\mathcal{M}}{{\mathcal{R}}^3}\right)N_{ab},\nonumber\\
Y_{ab}=\left[\dfrac{1}{2}(p+\rho)-\left(\Pi+\dfrac{2\mathcal{M}}{{\mathcal{R}}^3}\right)\right]e_{a}e_{b}+\dfrac{1}{2}\left[p+\left(\Pi+\dfrac{2\mathcal{M}}{{\mathcal{R}}^3}\right)\right]N_{ab}.
\end{align}
We can now transparently express the energy momentum tensor for a general matter field (in 1+1+2 decomposition) in terms of structure scalars  as

\begin{equation}
\left[T_{ab}\right]=\begin{pmatrix}
X_{T} &  \sqrt{2}Z &0&0\\
\sqrt{2}Z &P_{r}  &0&0\\
0&0& P_{\bot} & 0 \\
0&0&0& P_{\bot}
\end{pmatrix},
\end{equation}
with the following radial and tangential pressures (on the spherical 2-shell)
\begin{align}
P_{r}=\frac{2}{3}\big(Y_{T}-Y_{TF}-X_{TF}\big)-\frac{X_{T}}{3},\nonumber\\
P_{\bot}=\frac{1}{3}\big({2Y_{T}}+{Y_{TF}}-{X_{T}}+{X_{TF}}\big).
\end{align}
Consequently, we compute the Ricci tensor and the Ricci scalar in terms of these scalars as
\begin{align}
R_{ab}=&Y_{T}u_au_b+\left[\left(\dfrac{2X_{T}-Y_{T}}{3}\right)-\frac{2}{3}(X_{TF}+Y_{TF})\right]e_ae_b+\sqrt{2}Z(u_ae_b+u_be_a)\nonumber\\
&+\left[\left(\dfrac{2X_{T}-Y_{T}}{3}\right)+\dfrac{(X_{TF}+Y_{TF})}{3}\right]N_{ab},\,\,\,\,R_{a}{}^{a}=R=2(X_{T}-Y_{T}).
\end{align}

Finally, for a Type II matter field the structure scalars should satisfy the following relation
\begin{equation}\label{Typ2c}
\mid\sqrt{2}Z\mid=\frac{1}{3}(X_{T}+Y_{T})-\frac{1}{3}(X_{TF}+Y_{TF}).
\end{equation}
This can be derived from the condition of Type II matter field discussed previously as in Eq.(\ref{eigen2}).

\section{Generalized Vaidya: Type II matter field}
We consider a spherically symmetric non-static spacetime in Eddington-Bondi coordinates
\begin{equation}
ds^2=-A(v,r)^2f(v,r)dv^2+2\epsilon A(v,r) dvdr + r^2(d\theta^2+\sin^2\theta d\phi^2),\;\;\epsilon = \pm 1,
\end{equation}
which satisfies the Einstein field equations
\begin{equation}\label{Ein}
R_{ab}-\frac{1}{2}R g_{ab}= T_{ab}.
\end{equation}
We introduce a local mass function $m$ via $ f(v,r)=1-2m(v,r)/r $, related to the gravitational energy within a given radius $r$. Therefore, the  Misner-Sharp mass is also $\mathcal{M}=m(v,r)$.
We can now demand that the EMT $ T_{ab} $ satisfies the conditions $ T_{r}^{v}=0 $ and $ T_{\theta}^{\theta}=kT_{r}^{r} $ to obtain $ A(v,r)=g(v) $. Subsequently, we can always define another null coordinate $\bar{v}=\int g(v)dv$ and set $ A(v,r)=1 $, without any loss of generality. Finally, we obtain the familiar form of the generalized Vaidya spacetime \cite{WW,DG} 
\begin{equation}\label{genv}
ds^2=-\left(1-\frac{2m(v,r)}{r}\right)dv^2+2\epsilon dvdr + r^2(d\theta^2+sin^2\theta d\phi^2),
\end{equation}
and the EMT can be generally expressed in the following manner
\begin{equation}
T^{a}_{b}=\begin{pmatrix}
T^{v}_{v} &  0 &0&0\\
T^{r}_{v} & T^{r}_{r} &0&0\\
0&0& T^{\theta}_{\theta} & 0 \\
0&0&0& T^{\phi}_{\phi}
\end{pmatrix}.
\end{equation}
This EMT in general belongs to a Type II fluid with a double null eigenvector. When $\epsilon = + 1$, the null coordinate $v$ represents the Eddington advanced time, in which $r$ is decreasing towards the future  along a ray $v = \text{const.}$ (ingoing), while when $\epsilon = - 1$, it represents the Eddington retarded time, in which $r$ is increasing towards the future  along a ray $v = \text{const.}$ (outgoing).

Using Eq.(\ref{genv}) with the Einstein field equations $(\ref{Ein})$, the corresponding EMT can be written as
\begin{equation}
\label{eq3}
T_{ab} = T^{(n)}_{ab} + T^{(m)}_{ab},
\end{equation}
with
\begin{eqnarray}
\label{eq4}
T^{(n)}_{ab} &=& \tilde{\mu} l_{a}l_{b},\nonumber\\
T^{(m)}_{ab} &=& (\tilde{\rho} + \tilde{P}) \left(l_{a}n_{b} + 
l_{b}n_{a}\right) + \tilde{P} g_{ab},
\end{eqnarray}
where $T_{ab}^{(n)}$ is the pure null radiation component (Type II matter) that moves along the null hypersurfaces $v= \text{const.}$ 
If we take $\tilde{\rho} = \tilde{P} = 0$, the solutions reduce to the pure Vaidya solution with $m = m(v)$. Here $T_{ab}^{(m)}$ is the Type I component. Although, as a whole $T_{ab}$ is of Type II classification. 
Projecting the EMT of Eq.(\ref{eq3}) into the 1+1+2 form, spanned by $ \{u_{a}, e_{a}, N_{ab}\} $
\begin{equation}
\label{eq7}
l_{a} = \frac{u_{a} + e_{a}}{\sqrt{2}},\;\;
n_{a} = \frac{u_{a} - e_{a}}{\sqrt{2}},\;\;
N_{ab}=g_{ab}+u_{a}u_{b}-e_{a}e_{b}=g_{ab}+2l_{(a}n_{b)},
\end{equation}
we find that
\begin{equation}
\label{eq8}
T_{ab}=\left(\dfrac{\tilde{\mu}}{2}+\tilde{\rho}\right)u_{a}u_{b}+\left(\dfrac{\tilde{\mu}}{2}-\tilde{\rho}\right)e_{a}e_{b}+\dfrac{\tilde{\mu}}{2}\left(u_{a}e_{b}+u_{b}e_{a}\right)+\tilde{P}N_{ab},
\end{equation}
which in general belongs to the Type II fluids and has the form of Eq.(\ref{EM1}). The null vector $l^{a}$ is a double null eigenvector of the EMT, where $ \rho=\left({\tilde{\mu}}/{2}+\tilde{\rho}\right) $ is the energy density, the radial pressure $ P_{r}=\left({\tilde{\mu}}/{2}-\tilde{\rho}\right) $, the heat flux $ Q={\tilde{\mu}}/{2} $ and the tangential pressure $ P_{\bot}=\tilde{P} $. Hence, the 1+1+2 matter variables
$\{\rho, p, Q, \Pi\}$ are obtained as follows

\begin{align}\label{matter1}
\rho=& \left(\dfrac{\tilde{\mu}}{2}+\tilde{\rho}\right), \nonumber\\
Q=& \dfrac{\tilde{\mu}}{2},\nonumber \\
p=& \dfrac{1}{3}(P_{r}+2P_{\bot})=\dfrac{1}{3}\left(\dfrac{\tilde{\mu}}{2}-\tilde{\rho}+2\tilde{P}\right), \nonumber\\
\Pi=& \dfrac{2\tilde{\Pi}}{3}=\dfrac{2}{3}\left(\dfrac{\tilde{\mu}}{2}-\tilde{\rho}-\tilde{P}\right).
\end{align}

Thus, a timelike observer will encounter the matter fluid  with the energy density $ \rho $, the isotropic pressure $p$, the radial heat flux $Q $ and the trace-free part of the anisotropic pressure $\Pi$ with the values given in Eq.(\ref{matter1}). Consequently, the structure scalars can be readily found using Eq.(\ref{structure}) to give
\begin{eqnarray}
Y_T=\frac{1}{2}(\tilde{\mu}+2\tilde{P}) , \qquad
Y_{TF}=\dfrac{3}{2}\mathcal{E}+\frac{1}{2}\left(-\dfrac{\tilde{\mu}}{2} +\tilde{\rho}+\tilde{P}\right) ,\label{EY}
\\
X_T= \left(\dfrac{\tilde{\mu}}{2}+\tilde{\rho}\right) , \qquad
X_{TF}=-\dfrac{3}{2}\mathcal{E}+\frac{1}{2}\left( -\dfrac{\tilde{\mu}}{2} +\tilde{\rho}+\tilde{P}\right),\label{EX}
\\
\centering Z= \tilde{\mu}/2\sqrt{2},
\end{eqnarray}
where the electric part of the Weyl tensor $\mathcal{E}$ is determined ($\mathcal{R}=r$) by the following
\begin{equation}
\mathcal{E}=\dfrac{\rho}{3}-\dfrac{\Pi}{2}-\dfrac{2\mathcal{M}}{r^3}=-\dfrac{2\mathcal{M}}{r^3}+\dfrac{1}{3}(2\tilde{\rho}+\tilde{P}).
\end{equation}
Hence the energy density of Type II field component does not contribute to the Weyl tensor. Thus using Eq.(\ref{EY}) we obtain the complexity factor for the generalized Vaidya spacetime
\begin{equation}\label{complexity}
Y_{TF}=-\left(\dfrac{3\mathcal{M}}{r^3}+\frac{\tilde{\mu}}{4}\right)+\frac{3\tilde{\rho}}{2}+\tilde{P}.
\end{equation}
It is important to mention that the obtained forms of the structure scalars are covariant. For the null vectors (proposed in \cite{WW})
\begin{equation}
l_{a}=\delta_{a}^{0},\,\,\,\,\,n_{a}=\frac{1}{2}\left[1-\frac{2m(v,r)}{r}\right]\delta_{a}^{0}-\epsilon\delta_{a}^{1},\,\,\, l_{a}n^{a}=-1,
\end{equation}
the matter variables become
\begin{equation}
\label{eq5}
\tilde{\mu} = \frac{2 \epsilon \dot{m}(v, r)}{r^{2}},\;\;
\tilde{\rho}= \frac{ 2 m'(v, r)}{r^{2}},\;\;
\tilde{P}= - \frac{ m''(v, r)}{r},
\end{equation}
where
$$
\dot{m}(v, r) \equiv \frac{\partial m(v, r)}{\partial v}, \;\;\;\;\;\;
m'(v, r) \equiv \frac{\partial m(v, r)}{\partial r}.
$$ 

In particular, when $ m = m(v)$, as we mentioned previously, the solutions reduce to the Vaidya solution, and the energy conditions (weak, strong, and dominant) all reduce to $\tilde{\mu}\ge 0$. When $m = m(r)$, we have $\tilde{\mu} = 0$, and the matter field degenerates to Type I fluid \cite{WW}, i.e., the non-dissipative case corresponds to $ Z=0 $. In general, the complexity increases with the Weyl tensor. However, in the isotropic case $\Pi=0$, the complexity factor is proportional to the Weyl scalar, which is also the gravitational entropy \cite{CET}. From the obtained expression of the complexity factor it is clear that both the geometrical (Misner-Sharp mass) and the matter/thermodynamical contributions are at play. If we consider Penrose's Weyl curvature hypothesis, then the gravitational entropy is only one of the contributors in the measure of complexity. The other contribution is coming from the matter sector, specifically via the pressure anisotropy. However, the propagation and evolution of pressure anisotropy depends on the Weyl tensor and other quantities. In the subsequent section we will analyse specific examples to understand the complexity further.

\subsection{Some subclasses of generalized Vaidya }

\begin{itemize}
\item \textbf{Vanishing complexity:} From the above expression of complexity (\ref{complexity}), we want to explore which spacetime
corresponds to minimal complexity, i.e., which spacetime has vanishing complexity. We obtain the general condition for vanishing complexity as
\begin{equation}\label{zeroY}
\mathcal{M}=\frac{\mathcal{R}^3}{3}\left(2\rho+\frac{3p}{2}-3\vert Q\vert\right).
\end{equation}
Essentially, the geometrical part $\{\mathcal{M}, K\}$ should balance the matter part $\{\tilde{\mu}, \tilde{\rho}, \tilde{P}\}$, making the complexity vanish. It turns out if we make $ \tilde{\rho}=-\tilde{P}=\Lambda $ and $ \tilde{\mu}=0 $, we get 
\begin{equation}
Y_{TF}=0. 
\end{equation}
This corresponds to the known de Sitter spacetime with
\begin{equation}
\mathcal{M}=m(r)=\frac{\Lambda r^3}{6}.
\end{equation}
 If we do not take $ \Lambda $ into account, the simplest spacetime corresponds to the flat Minkowski spacetime, with $ \mathcal{M}=0 $.
Apart from the above mentioned trivial solution, we can compute the general condition (by substituting Eq.(\ref{eq5}) in Eq.(\ref{zeroY})) for vanishing complexity as the following
\begin{equation}
2m^{''}r^2-6m^{'}r+\epsilon\dot{m}r+6m=0
\end{equation}
with a general non-trivial solution 
\begin{align}
m(v,r)=& \frac{C_{1}e^{\beta v}}{\sqrt{\epsilon \beta}}\Bigg\{(C_{2}J_{1}(z)+C_{3}Y_{1}(z))\sqrt{2}r^{3/2} \nonumber\\
& -r^{2}\sqrt{\epsilon\beta}(C_{2}J_{0}(z)+C_{3}Y_{0}(z))\Bigg\},
\end{align}
where $ z\equiv\sqrt{2\epsilon \beta r} $. $J$ and $Y$ are the Bessel functions of the first and second kind respectively, with $\beta,C_{i}$s as constants of integration. In the limit $z<<1$ (small $r$), the dominant term in the mass function is linear in $r$ as $m\approx -2C_{1}C_{3}re^{\beta u}/\pi \epsilon \beta+\mathcal{O}(r^2 \ln{r})$ when $ C_{3}\neq 0 $. Hence, the $ m\approx \Phi(v) r $, where $ \Phi(v)\equiv 2\mid C_{1}C_{3}\mid e^{\beta v}/\pi \epsilon \beta $, as $ C_{1}C_{3}  $ needs to be negative for a positive mass. Therefore near the center of the system, it has a diverging matter density profile indicating a naked singularity, because in general $ g_{uu}\neq 0 $. Please note $ g_{uu} $ can be made to vanish for a special value $ v_{\ast}=\ln(\pi \epsilon \beta/4\mid C_{1}C_{3}\mid)/\beta $, however it is not a true horizon (valid for all $ r $) and also it fails to trap the non-radial light rays from escaping to infinity. Hence, $ C_{3}\neq 0 $ corresponds to naked singularity violating CCC. 

However, if we impose $ C_{3}=0 $, we obtain a surprising result for the small $ r $ limit 
\begin{equation}\label{regM}
m(v,r)\approx \frac{C_{1}C_{2}(\epsilon \beta)}{4} e^{\beta v} r^3,
\end{equation}
making the matter density finite positive and homogeneous near the central region. Therefore, it has no central singularity and $C_{ 
3}=0$ ensures initial regularity, allowing horizons to form dynamically during collapse. This adheres to CCC, as any singularity (e.g., from continued collapse) is hidden behind a horizon at $ r_{H}=2m(v,r_{H})>0 $. Additionally, if we impose the condition $ C_{1}C_{2}=2\Lambda/3\epsilon \beta $ in Eq.(\ref{regM}), we obtain the mass $ m(v,r)=e^{\beta v}\Lambda r^3/6 $, which is kind of an evolving de Sitter mass. However, the important finding is the existence of a regular homogeneous solution for $ C_{3}=0 $ having vanishing complexity. In general, $ \beta>0 $ makes the mass grow exponentially (accretion) and negative $\beta$ implies evaporation.

In the large distance limit, the general mass function is oscillatory in nature (superposition of ingoing/outgoing radiation) as in  
\begin{equation}
m(v,r)\approx -C_{1}e^{\beta v}\left(\frac{2r^7}{\pi^2 \epsilon \beta}\right)^{1/4}\left[C_{2}\cos(\theta)+C_{3}\sin(\theta)\right],
\end{equation}
where $\theta=z-\pi/4$. However, as we discussed previously, we can safely take $C_{3}=0$ (for regularity). Now, to have a positive mass in this limit $ \cos(\theta) $ has to be negative, which implies that, (in the asymptotic limit) there exist solutions of positive mass shells for different intervals of $ r $ (for $\epsilon \beta<<1$). Sticking to the first positive mass shell, the plausible domain of $ r $ can be truncated and matched with an external spacetime at $r=\mathcal{R}_{B}\in (9\pi^2/32\epsilon\beta,49\pi^2/32\epsilon\beta) $ or by simply restricting the solution to the core region $ r<<\mathcal{R}_{min}=9\pi^2/32\epsilon\beta $ (avoiding oscillations completely). It is clear that in general the system is not asymptotically flat (similar to the de Sitter case). It can always be adjusted by matching it with an external metric at a finite radius or embedding it (the solution is not isolated) in a dynamic background.

\item \textbf{Monopole:} If we now take the mass parameter as $ m(r)=ar/2 $, we obtain the thermodynamic variables as $ \tilde{\rho}=a/r^2, \tilde{\mu}=\tilde{P}=0$. Under such conditions the fluid reduces to a Type I matter field, and satisfies all the three energy conditions when $ a>0 $. It is again a well known solution which represents the gravitational field of a monopole and the corresponding complexity is given by the following
\begin{equation}
Y_{TF}=0.
\end{equation}
Therefore from the complexity perspective the monopole and de Sitter/Minkowski solutions are non different. To understand this phenomena further we now proceed to analyse other spacetimes.  

\item \textbf{Pure Vaidya:} As mentioned earlier we get the complexity of the pure Vaidya spacetime when we take $ \tilde{\rho}=\tilde{P}=0 $, which implies the mass function has no radial dependence $ m=f(v)$. The expression of $Y_{TF}$ becomes
\begin{equation}
Y_{TF}=-\left(\frac{3\mathcal{M}}{r^3}+\frac{\tilde{\mu}}{4}\right).
\end{equation}
All the three energy conditions (weak, strong and dominant) demand $ \tilde{\mu}\geq 0 $, making the complexity a negative quantity. This is a very interesting result for a pure Type II fluid. To understand it further we substitute the respective values of mass and energy density to get the following
\begin{equation}
Y_{TF}=-\frac{\epsilon}{2r^3}(6\epsilon f +r \dot{f}),\,\,\,\, \epsilon=\pm 1.
\end{equation}
For the sake of argument if we want $ Y_{TF} $ to be positive, then it implies $ (3f/r + \epsilon\dot{f}/2) $ has to be a negative quantity. For $ \epsilon=1 $, the only possible solution is when $ \dot{f}<0 $, making $ \tilde{\mu} $ a negative quantity. This violates the energy conditions. Further, the heat flux $ Q $ also becomes negative. This is situation where matter-energy flux is flowing inwards, however we have a mass loss. Clearly this cannot be a physical situation. Similarly, when $ \epsilon=-1 $, $ \dot{f} $ has to be negative following the same arguments. However, in this case a positive complexity factor is possible when when $ f $ becomes negative, making the mass function a negative quantity. Hence, the second situation is also not physical. Therefore, in pure Vaidya spacetime, a Type II fluid has to have negative complexity. It seems negative complexity factor means the system is purely Type II null radiation fluid. This might be an artefact of using a timelike observer for a null fluid. However, it is interesting to see that under the Herrera's scheme of complexity, pure Type II fluid has negative value favoured by all the three energy conditions.

\item \textbf{Charged Vaidya:} For the charged Vaidya solution the corresponding mass function is $m(v,r)=f(v)-{q^2(v)}/{2r}$, giving the energy density and pressure as $\tilde{\rho}=\tilde{P}={q^2}/{r^4}$, $\tilde{\mu}={2\epsilon}(r\dot{f}-q\dot{q})/{r^3}$. We can now easily obtain the corresponding complexity factor as the following
\begin{equation}
Y_{TF}=-\frac{\epsilon}{2r^3}(6\epsilon f +r \dot{f}-q \dot{q})+\frac{4q^2}{r^4},\,\,\,\, \epsilon=\pm 1.
\end{equation}
We can clearly observe the extra contributions due to charge in the complexity as opposed to the pure Vaidya case. The three energy conditions ensure that we are beyond the critical radius $ r>q\dot{q}/\dot{f} $ with a physically plausible condition $\dot{f}/\dot{q}>0$ and the term $ \epsilon(r\dot{f}-q\dot{q}) $ is always positive. The non-negativity of mass implies $ (f-q^2/2r) $ is also positive. Therefore, the complexity factor can now be rewritten in the following manner
\begin{equation}
Y_{TF}=\underbrace{\frac{5}{2}\frac{q^2}{r^4}}_\text{positive}+\overbrace{\frac{3}{r^3}\left(\frac{q^2}{2r}-f\right)-\frac{\epsilon}{2r^3}\left(r\dot{f}-q\dot{q}\right)}^\text{negative}.
\end{equation}
The above equation clearly demonstrates that the complexity has both positive and negative contributions. The complexity can be of any sign including zero depending upon which term dominates or balances each other. In the following we neatly present the signs of complexity with their corresponding conditions.
\begin{equation}
Y_{TF}\,\,\gtreqqless \,\, 0, \,\,\,\,\,\implies 3rf+\frac{\epsilon}{2}r^2\dot{f}\,\,\lesseqqgtr \,\, 4q^2+\frac{\epsilon}{2}rq\dot{q}.
\end{equation}
The introduction of charge makes the complexity of the system richer. If the charge terms dominate or equal to the "neutral terms" then we have positive or vanishing complexity. However, in a physically realistic system the charge is minute. Therefore the neutral terms dominate over the charge terms, giving rise to negative complexity factor.

\item \textbf{Husain solution:} Another known general solution is the Husain solution where the mass function is the following
\begin{equation}
m(v,r)=f(v)-\frac{g(v)}{(2k-1)r^{2k-1}}.
\end{equation} 
Therefore, the corresponding energy density is obtained as 
\begin{equation}
\tilde{\mu}=\frac{2\epsilon}{r^2}\left[\dot{f}(v)-\dfrac{\dot{g}(v)}{(2k-1)r^{2k-1}}\right]. 
\end{equation}
We also obtain the equation of state $ \tilde{P}=k\tilde{\rho}={2kg(v)}/{r^{2k+2}} $ in the Type I sector. Therefore, the complexity factor can be readily obtained as  
\begin{equation}
Y_{TF}=-\frac{\epsilon}{2r^3}\left[6\epsilon f+\dot{f} r -\dfrac{\dot{g}}{(2k-1)r^{2k-2}}\right]+\frac{4gk}{r^{2k+2}}\left[\frac{k+1}{2k-1}\right].
\end{equation}
The energy conditions demand that $ g $ and $ k $ need to be positive. Subsequently, we follow a similar analysis (as for charged Vaidya) and rewrite the $ Y_{TF} $ in the following manner
\small \begin{equation}
Y_{TF}=\overbrace{(2k+3)\dfrac{g}{r^{2k+2}}}^\text{positive}+\underbrace{\frac{3}{r^3}\left[\frac{g}{(2k-1)r^{2k-1}}-f\right]-\frac{\epsilon}{2r^2}\left[\dot{f}-\dfrac{\dot{g}}{(2k-1)r^{2k-1}}\right]}_\text{negative}.
\end{equation}
It is evident that the $ Y_{TF}$ has significant negative contributions, and if it dominates over the positive term, the complexity becomes negative. We can always find out the condition of vanishing complexity by equating the two terms. Therefore the following condition holds
\begin{equation}
Y_{TF}\gtreqqless 0, \implies \frac{4k(k+1)}{2k-1}g+\dfrac{\epsilon \dot{g}r}{2(2k-1)}\gtreqqless r^{2k}\left(\frac{3f}{r}+\frac{\epsilon \dot{f}}{2}\right).
\end{equation}
Here, again we can notice the complexity deviates from the negative value when the Type I matter is present with the `$ g $ terms'. If the terms containing $ g $ contribute minimally, we have a negative $ Y_{TF} $. Now if we take $ g(v)={(2k-1)q^{2k}(v)}/{2} $, the Husain solution complexity corresponds to the charged Vaidya case for $ k=1 $.

\item \textbf{Monopole de Sitter charged Vaidya:}
As the matter variables are linear in terms of derivatives of the mass function, the superposition of previous cases is also a solution of the field equation. Therefore, the mass function is of the following form
\begin{equation}
m(v,r)=\frac{ar}{2}+\frac{\Lambda r^3}{6}+f(v)-\frac{q^2(v)}{2r}.
\end{equation}
The complexity factor then becomes
\begin{equation}
Y_{TF}=-\frac{\epsilon}{2r^3}(6\epsilon f +r \dot{f}-q \dot{q})+\frac{4q^2}{r^4},\,\,\,\, \epsilon=\pm 1.
\end{equation}
It is evident that the resulting complexity is the sum of monopole and charged Vaidya $ Y_{TF} $ and is independent of the cosmological constant $ \Lambda $, similar to the de Sitter case. It is interesting to see that the complexity factor is identical to the charged Vaidya case (monopole and cosmological constant does not contribute) having both positive and negative contributions as discussed previously.

\end{itemize}

{\renewcommand{\arraystretch}{2.5}
\begin{table}[hbt!]
   \large
   \centering
   \rotatebox{0}{
   \begin{minipage}{\textwidth}
   \centering
   \caption{Complexity of different subclasses of generalised Vaidya spacetime. } 
    \label{tabf}
    \resizebox{\textwidth}{!}{
   \begin{tabular}{lccr}
   \toprule\toprule
  {\large\textbf{Spacetime}} & {\large\textbf{Mass function}} & {\large\textbf{Matter variables}} & {\large\textbf{Complexity} }\\ 
   \midrule
  de Sitter & $m(r)=\frac{\Lambda r^3}{6}$ & $ \tilde{\mu}=0,\,\, \tilde{\rho}=-\tilde{P}=\Lambda $ & $ Y_{TF}=0$ \\
   Monopole & $m(r)=\frac{ar}{2} $  & $\tilde{\mu}=\tilde{P}=0,\,\, \tilde{\rho}=\frac{a}{r^2} $ & $ Y_{TF}=0 $\\
 Pure Vaidya &  $m(v)=f(v)$ & $\tilde{\rho}=\tilde{P}=0  $ & $Y_{TF}=-\frac{\epsilon}{2r^3}(6\epsilon f +r \dot{f}) $\\   
 Charged Vaidya &  $m(v,r)=f(v)-\frac{q^2(v)}{2r}$ & $\tilde{\rho}=\tilde{P}=\frac{q^2}{r^4},\,\, \tilde{\mu}=\frac{2\epsilon}{r^3}(r\dot{f}-q\dot{q})$ & $Y_{TF}=-\frac{\epsilon}{2r^3}(6\epsilon f +r \dot{f}-q \dot{q})+\frac{4q^2}{r^4}$\\  
 Husain solution & $m(v,r)=f(v)-\frac{g(v)}{(2k-1)r^{2k-1}}$ & 
$\!\begin{aligned}[t] 
&  \tilde{P}=k\tilde{\rho}=\dfrac{2kg(v)}{r^{2k+2}}, \\
& \tilde{\mu}=\frac{2\epsilon}{r^2}\left[\dot{f}(v)-\dfrac{\dot{g}(v)}{(2k-1)r^{2k-1}}\right]  
\end{aligned}$ & 
$\!\begin{aligned}[t] 
 Y_{TF}=& -\frac{\epsilon}{2r^3}\left[6\epsilon f+\dot{f} r -\dfrac{\dot{g}}{(2k-1)r^{2k-2}}\right]  \\
& +\frac{4gk}{r^{2k+2}}\left[\frac{k+1}{2k-1}\right] 
\end{aligned}$\\ 
$\!\begin{aligned}[t] 
 & \text{Monopole dS charged} \\
 & \text{Vaidya} \end{aligned}$
 & 
$\!\begin{aligned}[t]  
   m(v,r)=&\frac{ar}{2}+\frac{\Lambda r^3}{6}\\
  & +f(v)-\frac{q^2(v)}{2r}\end{aligned}$  &
 $ \!\begin{aligned}[t]
\tilde{\mu}=\frac{2\epsilon}{r^3}(r\dot{f}-q\dot{q}),\,\,\,\, & \tilde{\rho}=\frac{a}{r^2}+\Lambda+\frac{q^2}{r^4},\\
&
\tilde{P}=-\Lambda +\frac{q^2}{r^4} \end{aligned}$ &  $Y_{TF}=-\frac{\epsilon}{2r^3}(6\epsilon f +r \dot{f}-q \dot{q})+\frac{4q^2}{r^4}$\\
   \bottomrule\bottomrule
   \end{tabular}
   }
   \end{minipage}}
\end{table}
}
All the above mentioned mass functions can be obtained by using a linear equation of state of the Type I matter component. Such well known spacetimes and their corresponding complexity are tabulated in Table \ref{tabf}. Similarly a quadratic and polytropic equation of state can be considered and the corresponding complexities can be readily obtained. For more details on equation of states and different solutions see \cite{BY,MGM}.

\textbf{Condition of negative complexity:}
For Type II fluids, the situation is different from Type I, because the stress tensor is fundamentally radiative with double null eigenvectors (no timelike eigenvector) making the complexity factor behave differently. In general the three energy conditions (weak, strong and dominant) impose the restrictions on the matter variables like 
\begin{equation}
\tilde{\mu} > 0, \,\,\,\,\,\, \tilde{\rho}>\tilde{P} > 0.
\end{equation}
Now using the previously obtained expression of complexity, we can get the following relation if the $ Y_{TF} $ is to be negative
\begin{equation}
\left(\frac{3\mathcal{M}}{r^3}+\frac{\tilde{\mu}}{4}\right)>\left(\frac{3\tilde{\rho}}{2}+\tilde{P}\right).
\end{equation}
This means that the pure Type II matter (null dust) and the geometrical terms should dominate over the Type I matter components. The trivial case would be the Schwarzschild metric (purely governed by geometric terms), where all the matter variables are zero and the complexity becomes $ Y_{TF}=-3\mathcal{M}/r^3 $ (a negative quantity). However, the negative complexity in Vaidya spacetime is a non-trivial result. Therefore, both static and radiating black holes are of negative complexity. In fact, in spherical symmetry $ Y_{TF}=-3\mathcal{M}_{schw}/r^3 $ corresponds to the simplest massive object possible (indicating a deeper meaning of the negative sign). Hence, we would argue that the negative sign of the complexity factor is important, indicating its origin as geometry and pure null fluid (at least for Type II matter), unlike Type I, where $ \Pi $ can also contribute negatively. This happens because in Type II matter, $\Pi$ is constrained by $ Q $ in (\ref{eigen2}), making the complexity as 
\begin{equation}
Y_{TF}=\left(2\rho+\frac{3p}{2}\right)-\frac{3\mathcal{M}}{\mathcal{R}^3}-3\vert Q \vert.
\end{equation}

\subsection{Evolution/Propagation equations of structure scalars}
In the previous section we obtained the covariant expression of complexity and other structure scalars for the generalized Vaidya spacetime. Next, we present how their directional derivatives change through the kinematical variables. Therefore, we compute the propagation and evolution equations of the structure scalars in a general LRS II spacetime. To simplify our calculations we will adopt the `static' $ \dot{K}=0 $ frame (we can always pick such an observer)\cite{GBCAC}. We take the $l_{a}$ and $n_{a}$ being two orthogonal null vectors so that for this observer the Gaussian curvature $K$ of the $2$-sheet is constant in time (but non-vanishing). 
\begin{eqnarray}
\label{eq6}
l_{a} &=& -\frac{1}{\sqrt{2}}\left[1 - \frac{2m(v, r)}{r}\right]^{\frac{1}{2}}\delta^{0}_{a},\nonumber\\
n_{a} &=& -\frac{1}{\sqrt{2}}\left[1 - \frac{2m(v, r)}{r}\right]^{\frac{1}{2}}\delta^{0}_{a} + \sqrt{2}\epsilon\left[1 - \frac{2m(v, r)}{r}\right]^{-\frac{1}{2}} \delta^{1}_{a},\nonumber\\
l_{a}l^{a} &=& n_{a}n^{a} = 0, \;\;
l_{a}n^{a} = - 1.
\end{eqnarray}
In this frame the shear $ \Sigma $ is proportional to the expansion scalar $ \Theta $, i.e., $ \Sigma=2\Theta/3 $, since $ \dot{K}=(\Sigma-2\Theta/3)K=0 $. Therefore, in $1+1+2$ covariant decomposition the equations become

\begin{itemize}
\item \it{Propagation:}
\begin{align}
&6\hat{\phi}+3\phi^2=4(-X_{T}+X_{TF}),\\
& \sqrt{2}Z=-\frac{3}{2}\phi\Sigma,\\
& \hat{X}_{TF}+\frac{\hat{X}_{T}}{2}=-\frac{3}{2}\phi X_{TF}.
\end{align}
\item \it{Evolution:}
\begin{align}
&\dot{\phi}=\sqrt{2}Z, \\
& 3\mathcal{A}\phi=2(Y_{T}-Y_{TF}), \\
& \dot{X}_{TF}+\frac{\dot{X}_{T}}{2}=-\frac{3}{2\sqrt{2}}\phi Z. \label{Tdot}
\end{align}
\item \it{Propagation/evolution:}
\begin{align}
&\hat{\mathcal{A}}-\dot{\Theta}=-(\mathcal{A}+\phi)\mathcal{A}+\Theta^2+Y_{T},\\
& \dot{X}_{T}+\sqrt{2}\hat{Z}=-\dfrac{2\Theta}{3}\left[(X_{T}+Y_{T})-(X_{TF}+Y_{TF})\right]-(\phi+2\mathcal{A})\sqrt{2}Z,
\end{align}
\begin{align}
  6\sqrt{2}\dot{Z}-4\left(\hat{X}_{TF}+\frac{\hat{X}_{T}}{2}\right)
 +4(\hat{Y}_{T}-\hat{Y}_{TF})=&\left(6\phi+4\mathcal{A}\right)(X_{TF}+Y_{TF}) \nonumber\\
 &-4\mathcal{A}(X_{T}+Y_{T})-12\sqrt{2}\Theta Z.
\end{align}
\end{itemize}

The above equations clearly demonstrate how the kinematical variables $\{\mathcal{A}, \phi, \Theta, \Sigma\}$ influence the propagation and evolution of these structure scalars. By substituting the covariant scalars we can get the propagation/evolution equations of structure scalars for any spherically symmetric spacetime. For Type II matter field we impose the additional condition from Eq.(\ref{Typ2c}).
\subsection{Gaussian curvature and Shell-complexity}
We will now make use of the 2-shell complexity introduced in \cite{SS}. The authors showed that there is a way to define complexity via the information in the change of Gaussian curvature of a given 2-shell. Interestingly, this gives rise to a wave equation (hyperbolic), which shows the causal behaviour determined by the second order source-free differential equation. To obtain the wave equation for the Gaussian curvature the authors introduce two new variables $ \Psi $ and $X$ defined as 
\begin{align}
\Psi \equiv&\,\, \Sigma - \frac{2}{3}\Theta, \label{Psi} \\
X \equiv&\,\, \mathcal{E}-\frac{1}{3}\mu+\frac{1}{2}\Pi=-\frac{2}{3}\left(X_{TF}+\frac{X_{T}}{2}\right).
\end{align}
We get the required closed form wave equation
\begin{equation}
\ddot K - \hat{\hat K} +\mathcal{F} K =0\,,
\end{equation}   
where the term `$\mathcal F$' behaves like the {\em restoration factor} of a travelling disturbance in an elastic medium with the general form as
\begin{equation}
\mathcal{F}=\left[6K-16\mathcal{M} K^{3/2}+\Psi^2 +\Psi\Theta+\mathcal{A}\phi+\mu-p-\Pi\right].
\label{F2}
\end{equation}
 It is interesting to see that in the most general scenario, the restoration factor depends (either explicitly, or via the equation of state) on all the geometrical and thermodynamical variables, which form the set
\begin{equation}
 \mathcal{D}\equiv\{\mathcal{A}, \Theta, \phi, \Sigma, \mathcal{E}, \mu, p, \Pi, Q\}\,.
\end{equation}
In this prescription the number of independent 1+1+2 geometrical and thermodynamical variables that the restoration factor depends on, gives a measure of the complexity of the spherical self-gravitating system.
In the case of generalised Vaidya spacetime the causality equation of the Gaussian curvature is simply 
\begin{equation}
 \hat{\hat K} =\mathcal{F} K.
\end{equation}
This can be obtained when $ \Psi=0 $ (corresponds to $\dot{K}=0$), which means the concerned quantities are $\{\Theta, \phi, \mathcal{E}, \mu, p, \Pi\} $, with an implicit dependence on $ Q $ via the equation of state of Type I matter field. It is interesting to observe how the restoration factor depends on the structure scalars
\begin{equation}
\mathcal{F}=\frac{2}{3}(X_{T}-X_{TF})+\frac{3}{2}\phi^2.
\end{equation}
Using the evolution equation (\ref{Tdot}) we can rewrite $\mathcal{F}$ as 
\begin{equation}
\mathcal{F}=\frac{4}{3Z^2}\left(\dot{X}_{TF}+\frac{\dot{X}_{T}}{2}\right)^2+\frac{2}{3}(X_{T}-X_{TF}).
\end{equation}
Now from the previously obtained propagation/evolution equations, $ Z $ can be completely expressed in terms of $ \{X_{T},X_{TF}\} $ and their directional derivatives in the following manner
\begin{equation}
Z=\frac{1}{\sqrt{2}}\left(\frac{\dot{T}}{\hat{T}}\right)(2T-X_{T}),\,\,\,\,\, T\equiv X_{TF}+\frac{X_{T}}{2}.
\end{equation}
Here $ T $ can be interpreted as "Misner-Sharp mass density" as
$T=3\mathcal{M}K^{3/2}$. Therefore, the wave equation of the Gaussian curvature takes the form
\begin{equation}
\hat{\hat{K}}=\frac{2}{3}\Bigg\{\big(X_{T}-X_{TF}\big)+\left(\frac{\hat{T}}{X_{TF}}\right)^2\Bigg\}K.
\end{equation}
Therefore the structure scalars $X_{T}$ and $ X_{TF} $ and their directional derivatives (related through $Z$) are explicitly driving the Gaussian curvature wave equation. The complexity factor $ Y_{TF} $ and the scalar $ Y_{T} $ influence the wave equation (implicitly) via the propagation and evolution equations. Therefore, the role of complexity on the Gaussian curvature of spherical shells is rather indirect.

Another important quantity on the shell is the Misner-Sharp mass, describing the total mass-energy inside the shell. The derivatives of the  Misner-Sharp mass are also related to the scalars $\{X_{T},X_{TF}\}$ by the following
\begin{equation}
\dot{\mathcal{M}}=\frac{\dot{T}}{3K^{3/2}},\,\,\,\,\, \hat{\mathcal{M}}=\frac{X_{T}\hat{T}}{3K^{3/2}(X_{T}-2T)},
\end{equation}
and  are related to each other via the structure scalars $ X_{T} $ and $ Z $ 
\begin{equation}
\frac{\hat{\mathcal{M}}}{X_{T}}+\frac{\dot{\mathcal{M}}}{\sqrt{2}Z}=0,
\end{equation}
where $X_{T}$ is the energy density and $Z$ is the heat flux. 
Hence, $\mathcal{M}$ can be uniquely determined by $\{X_{T},Z\}$, subject to required boundary conditions. Taking the second derivatives and using the commutator relation from Eq.(\ref{Comm}) we can get the following second order wave equation for $ \mathcal{M} $
\begin{align}
\frac{\sqrt{2}Z}{X_{T}}\hat{\hat{\mathcal{M}}}-\frac{X_{T}}{\sqrt{2}Z}{\ddot{\mathcal{M}}}+&\left(\Theta+\frac{\dot{X_{T}}}{X_{T}}-\frac{\sqrt{2}Z}{X_{T}}\frac{\hat{X_{T}}}{X_{T}}\right)\hat{\mathcal{M}} \nonumber\\
& +\left(-\mathcal{A}+\frac{X_{T}}{\sqrt{2}Z}\frac{\dot{Z}}{Z}-\frac{\hat{Z}}{Z}\right)\dot{\mathcal{M}}=0.
\end{align}
From the above relations it is evident that the $\mathcal{M}$ and it's propagation/evolution are explicitly governed by the $X_{T}$ and $Z$ scalars (with their directional derivatives), depicting how the Misner-Sharp mass changes through the energy density and heat flux. For Type II matter $ Z $ is also constrained by Eq.(\ref{Typ2c}).
\section{Concluding remarks}

In this paper we obtained the structure scalars (including the complexity factor) for Type II matter field and computed their covariant evolution/propagation equations in terms of $1+1+2$ variables. We found that for vanishing complexity there exists a class of non-trivial generalized Vaidya spacetimes with the following mass parameter
\begin{equation}
m(v,r)=\frac{r^3}{3}\Bigg\{-\frac{\tilde{\mu}}{4}+\frac{3\tilde{\rho}}{2}+\tilde{P}\Bigg\},
\end{equation}
apart from the trivial de Sitter and monopole solution. Interestingly the mass function of such solutions are proportional to $ r^{3} $ within appropriate limits, almost like a evolving de Sitter spacetime. We extended Herrera's complexity paradigm to null dust type matter fields and analysed the consequences. The obtained results are exact mathematical relations and are universally valid for any spherically symmetric system. The complexity factor is not a direct physical observable but depends on the density inhomogeneity (via Weyl scalar) and pressure anisotropy. A radiating star with pure Vaidya exterior is an excellent (complex) system to study it further observationally. The junction conditions relate the Type II field variables to that of the interior parameters (Type I). Consequently, one can employ standard astrophysical techniques and use the mass and radius of the known object to get a rough estimation of the complexity factor. Similarly one can employ the obtained (vanishing complexity) mass function to study astrophysical systems (BH shadows, gravitational waves, etc.). This line of inquiry may offer new insights in future. We have already presented a detailed mathematical analysis of the solutions.

 Another important finding of this paper is the physical meaning of negative complexity. In the Vaidya spacetime with pure Type II matter field (null radiation), the complexity is found to be always negative. Any extra component of Type I matter in the generalized Vaidya solution will contribute positively in the complexity, making the system more complicated and can take any sign. However, if the composite matter field is null radiation dominated, it will always have a negative complexity. Geometry also plays an important part in the sign of the complexity. We show transparently how our obtained expression of complexity for generalized Vaidya spacetime is physically consistent and can be employed to obtain different well known solutions with different matter fields creating a hierarchy of complexity. The above findings make the complexity of Type II matter field non-trivially different from that of Type I. 
 
It is important to mention that one should be careful about the Type III and Type IV matter fields, as they violate the energy conditions. The application of Herrera's structure scalars for these matter fields are not like Type I and Type II. It needs careful analysis, as they have Jordan block structure and complex eigenvalues respectively. The negative $\Delta$ may correspond to the Type IV and the decomposition can be done accordingly (needs a detailed study). However, Type III matter field inherently couples the preferred spatial direction of the 1+1+2 split to the transverse 2-sheet, violating the formalism's core assumption and therefore, the analysis needs a full complex null tetrad structure, like the Newman-Penrose formalism.

 We also investigated the influence of the structure scalars on the Gaussian curvature of the $2-$shell. Interestingly, the propagation/evolution of the Gaussian curvature is not explicitly dependent on the electric part of the Riemann tensor. The role of the complexity factor on the wave equation of the Gaussian curvature is indirect and is explicitly shown in the propagation/evolution equations. We further examined how the derivatives of Misner-sharp mass on the $2-$shell are dependent on the structure scalars $\{X_{T},Z\}$. Finally we obtain the wave equation of $ \mathcal{M} $ governed by $\{X_{T},Z\}$ and their directional derivatives. It is interesting to observe that there are no direct/explicit influence of $ Y_{TF} $ on these geometric equations of $ K $ and $ \mathcal{M} $.

From the expression of the complexity factor we can see that at each point it is determined by the following covariant scalars $\{\rho, p, \mathcal{M}, K, Q\}$. However, to gauge its behaviour we also need to take into account the evolution/propagation equations of the complexity factor, along with other structure scalars. Moreover, to understand the underlying physics in the complexity factor we need to analyse its different components and their origins. The super-energy $ W $ associated to the free gravitational field is defined using the Bel-Robinson tensor $ T_{abcd} $. In Petrov type D spacetimes we can directly obtain $W$ in terms of the Weyl scalar
\begin{equation}
W=T_{abcd}u^{a}u^{b}u^{c}u^{d}= \frac{1}{4}E^{ab}E_{ab}=\frac{3}{8}\mathcal{E}^2.
\end{equation}
Consequently, this super-energy $W$ is used to define (as free gravitational energy density) the CET gravitational entropy \cite{CET} (due to Weyl curvature hypothesis). Therefore, (in general) the complexity can be obtained in terms of the magnitude of  free gravitational field energy density $\rho_{grav}=\alpha\mid\mathcal{E}\mid$ and conventional mass-energy density
\begin{equation}
 Y_{TF}=3\mathcal{E}+\frac{3\mathcal{M}}{\mathcal{R}^3}-\frac{1}{2}X_{T}=Y_{TF}^{(grav)}+Y_{TF}^{(m)},
\end{equation}
where $ \mathcal{M}/\mathcal{R}^3 $ (mixed energy) has both the free gravitational field energy component via the Weyl scalar $ \mathcal{E} $ and the conventional matter (thermodynamic) energy components through matter energy density and pressure anisotropy. Whereas, $X_{T}$ is purely the matter energy density part. Therefore, the complexity has two distinct components, one due to free gravity $Y_{TF}^{(grav)}$ and another is from the matter fields $Y_{TF}^{(m)}$. For the generalised Vaidya spacetime, $ Y_{TF}^{(grav)}=3\mathcal{E}/2 $ and $Y_{TF}^{(m)}=Y_{TF}^{I}+Y_{TF}^{II}  $ has two components due to Type I and Type II matter fields with $Y_{TF}^{I}=(\tilde{\rho}+\tilde{P})/2  $ and $ Y_{TF}^{II}=-\tilde{\mu}/4 $. In pure Vaidya spacetime $Y_{TF}^{(grav)}$ component contributes negatively as $ \mathcal{E}=-2m(v)/r^3 $. Therefore, this transparently shows that the origin of the negative sign are the free gravity (geometry) and the Type II null component. Therefore, the (magnitude) component of the complexity due to free gravity field $ Y_{TF}^{(grav)} $ increases if the free gravitational energy increases. The gravitational entropy also increases with the structure formation. Therefore, the gravitational component of the complexity should increase with the structure formation.

 Hence, at least in pure vacuum the complexity (only gravitational component remains) seems to increase with the gravitational arrow of time. Future studies on the accreting/collapsing scenario with matching to external spacetimes should reveal more details.

\section*{Acknowledgements}
The authors thank the anonymous reviewers for their constructive feedback. SC is thankful to the University of KwaZulu-Natal (UKZN) for post doctoral funding. Special thanks to Prof. Sarbari Guha for her insightful comments. RG thanks UKZN and NRF, South Africa, for support. 
SDM thanks UKZN for research support.
\section*{Conflict of interest}
We confirm no conflict of interest exist.
\section*{Data availability statement}
All data that support the findings of this study are included within the article (and any supplementary files).

\end{document}